# Quantum-Ready Secure WAN:
# A Risk Assessment and Migration Framework

Saeed Alam

The university of Southern Mississippi

## Abstract

Enterprise wide-area networks (WANs) use quantum-vulnerable public-key cryptography to authenticate peers and establish keys for Internet Protocol Security (IPsec), Transport Layer Security (TLS), and software-defined WAN services. Harvest-now, decrypt-later collection already threatens data whose protection lifetime may exceed the time needed to migrate, while a future cryptographically relevant quantum computer would also endanger certificates and other authentication dependencies. Post-quantum standards and government directives now define algorithms and transition milestones, but they do not tell an enterprise which WAN services to migrate first or what evidence is sufficient for deployment.

This paper proposes a vendor-neutral framework that separates exposure priority, migration readiness, and evidence confidence, and maps those outputs to staged security, operational, and governance gates. It also defines a complete evaluation design for classical and ML-KEM hybrid key establishment in TLS 1.3 and IKEv2/IPsec. The retained artifact is narrower: it contains author-reported aggregate TLS p50, p95, and p99 latencies at configured loss settings of 0%, 1%, 3%, and 5%, an arithmetic checker, and limited environment and MTU/MSS notes. At those settings, the TLS-H minus TLS-C p95 differences are 0.00, 0.16, 0.38, and 0.63 ms, corresponding to 0.00%, 1.32%, 3.07%, and 5.00% relative to TLS-C. These values are descriptive aggregate observations, not inferential results, because per-trial data, sample counts, original captures, and timestamped logs were not retained. Protocol-size and MTU/MSS calculations are reported separately from measured behavior, and unsupported IKE, resource, throughput, retransmission, and downgrade outcomes are excluded.

**Keywords:** *post-quantum cryptography; quantum risk; enterprise WAN; cryptographic migration; cryptographic agility; ML-KEM; IPsec; TLS 1.3*

## 1 Introduction

Enterprise wide-area networks (WANs) carry business traffic between headquarters, branch offices, data centers, cloud environments, remote users, and external partners. These connections commonly use IPsec/IKEv2, TLS, and related public-key mechanisms to authenticate endpoints and establish symmetric session keys. Their security depends heavily on RSA, finite-field Diffie-Hellman, and elliptic-curve cryptography. These algorithms remain suitable against known classical attacks when correctly configured. However, a sufficiently capable quantum computer could solve the integer-factorization and discrete-logarithm problems on which their security depends [1]. Such a system - generally described as a cryptographically relevant quantum computer (CRQC) - is not known to exist, and its arrival date remains uncertain. The uncertainty does not remove the risk because cryptographic migration and data exposure operate on longer timelines.

An adversary does not need a CRQC today to create future harm. In a harvest-now, decrypt-later attack, encrypted traffic is collected and retained until a quantum computer or another cryptanalytic advance makes decryption practical. The immediate risk therefore depends not only on when a CRQC becomes available, but also on how long the intercepted information must remain confidential and how long the organization will need to replace vulnerable cryptography. If the required confidentiality period and migration time together extend beyond the estimated time to a CRQC, delayed migration can expose data before replacement is completed [2]. This condition is especially relevant to WAN traffic containing intellectual property, regulated personal information, authentication material, financial records, strategic plans, or government-controlled data.

The quantum threat also affects authentication and integrity. A future attacker capable of breaking RSA or elliptic-curve signatures could impersonate network peers, forge credentials, or undermine trust relationships that remain dependent on vulnerable public keys. Confidentiality exposure is immediate where recorded traffic has long-term value, whereas signature and authentication risks become acute when a CRQC is available. These two risk paths require different priorities and controls. Treating quantum readiness only as an encryption upgrade can therefore leave important dependencies - such as certificates, device identities, management interfaces, software validation, and routing-control relationships - outside the migration plan.

Post-quantum cryptography (PQC) provides algorithms designed to resist attacks by both classical and quantum computers. In 2024, the U.S. National Institute of Standards and Technology standardized ML-KEM for key establishment in FIPS 203, ML-DSA for digital signatures in FIPS 204, and SLH-DSA as a hash-based signature alternative in FIPS 205 [3-5]. These standards provide a stable technical basis for migration, but standardization does not by itself make an enterprise network quantum-ready. An organization must still discover where cryptography is used, identify protocol and certificate dependencies, establish migration priorities, determine product support, test interoperability, manage fallback behavior, and verify that new mechanisms operate acceptably across real network paths.

Government policy has made these planning requirements more concrete. OMB Memorandum M-23-02 requires U.S. federal agencies to maintain prioritized inventories of quantum-vulnerable cryptographic systems and consider system impact and data-protection lifetime [6]. Executive Order 14412 and OMB Memorandum M-26-15 accelerate the transition for high-value assets and high-impact systems: key establishment is targeted by December 31, 2030, digital signatures by December 31, 2031, and agency migration plans are due within 120 days of M-26-15 [7, 37]. These instruments directly govern federal agencies, but their inventory, prioritization, ownership, testing, and evidence requirements are useful reference points for private enterprises.

WAN migration presents problems that are not captured by algorithm benchmarks alone. Enterprise networks are heterogeneous and may contain physical routers, virtual appliances, firewalls, cloud gateways, remote-access clients, embedded devices, certificate services, and managed network services supplied by different providers. A tunnel can be technically capable of negotiating a post-quantum algorithm while remaining operationally unsuitable because of restricted path maximum transmission units, packet loss, added handshake messages, unsupported peers, or insecure fallback behavior. Migration may also cross organizational boundaries where neither party controls the complete cryptographic configuration.

These concerns are significant for post-quantum key establishment because keys and ciphertexts are generally larger than their classical counterparts. Current Internet Engineering Task Force (IETF) work integrates ML-KEM into TLS 1.3 and IKEv2, including hybrid exchanges that combine post-quantum and traditional shared secrets [8, 9]. For IKEv2, ML-KEM-768 uses an encapsulation key of 1,184 bytes and a ciphertext of 1,088 bytes, while the corresponding ML-KEM-1024 values are 1,568 bytes each. Protocol overhead can therefore cause messages to approach or exceed common path maximum transmission unit (MTU) limits. The IKEv2 specification work identifies fragmentation and downgrade behavior as operational security concerns [9]. A deployment decision must therefore consider protocol behavior and path constraints as well as cryptographic strength.

Prior research has established that post-quantum migration is an organizational and systems-engineering problem rather than a direct substitution of algorithms [2, 10]. Existing work has examined cryptographic inventories, dependency analysis, crypto-agility, algorithm performance, and post-quantum integration into individual protocols. These contributions provide important foundations. However, a practical gap remains between general migration guidance and the decisions faced by enterprise WAN operators. Available approaches do not fully integrate data lifetime, cryptographic exposure, operational criticality, migration effort, protocol dependencies, interoperability, constrained-path behavior, and deployment acceptance criteria into one WAN-specific assessment process. As a result, an enterprise may know that migration is necessary without having a defensible method for deciding which connection should be migrated first, what evidence is required before deployment, or how residual exposure should be documented.

This paper addresses that gap through the following research questions:

**RQ1.** Which security, data, operational, and implementation factors provide a defensible basis for prioritizing post-quantum migration across enterprise WAN services?

RQ2. What protocol-size implications and aggregate TLS latency differences can be established from the retained evidence, and what additional evidence is required for broader TLS and IKEv2 performance conclusions?

RQ3. How should measured or retained evidence be converted into migration waves, deployment acceptance thresholds, and residual-exposure decisions across heterogeneous environments?

To answer these questions, this paper makes four contributions.

1. It defines a WAN-specific threat model that distinguishes present harvest-now, decrypt-later exposure from future authentication, impersonation, and downgrade threats.
2. It develops a vendor-neutral prioritization model that keeps timing urgency, confidentiality and authentication impact, cryptographic control state, path exposure, and dependency concentration separate from migration readiness and evidence confidence.
3. It proposes a staged migration framework that maps exposure-priority bands and readiness states to controlled architecture, validation, pilot, deployment, retirement, and reassessment decisions.
4. It defines a reproducible full-evaluation method for classical and ML-KEM-based hybrid key establishment in IPsec/IKEv2 and TLS 1.3, and it demonstrates a strict evidence boundary using the limited retained TLS aggregate summary. The full method covers establishment latency, handshake size, processor and memory use, throughput, fragmentation, rekey, failure, and downgrade resistance; only the retained and arithmetic-checked quantities are presented as current results.

The intended output is an auditable decision process rather than a universal migration date or product recommendation. It helps an enterprise identify high-priority WAN services, select proportionate controls, establish procurement and interoperability requirements, and decide whether a deployment has satisfied predefined gates. This research-to-practice focus fits ACM DTRAP's emphasis on concrete digital threats, operational mitigation, and reproducible evidence [16].

The remainder of this paper is organized as follows. Section 2 presents the technical background, threat landscape, standards, and migration directives. Section 3 reviews related research and formalizes the research gap. Section 4 defines the enterprise WAN threat model. Section 5 introduces the quantum-exposure prioritization model, and Section 6 presents the migration framework. Section 7 specifies the complete experimental design and distinguishes it from the limited retained evidence. Section 8 reports the supported protocol-derived calculations and retained aggregate TLS summary. Section 9 provides a synthetic worked example. Section 10 discusses practical implications, limitations, and threats to validity, and Section 11 concludes the paper.

# 2 Background and Current Threat Landscape

## 2.1 Quantum Threat to Deployed Cryptography

Enterprise security protocols use two broad classes of cryptography. Symmetric algorithms, such as the Advanced Encryption Standard (AES), use the same secret key for encryption and decryption. Public-key algorithms use separate public and private keys and support functions such as key establishment, digital signatures, and endpoint authentication. IPsec and TLS normally use public-key cryptography during session establishment and symmetric cryptography to protect the subsequent data flow.
The distinction matters because quantum computing does not affect all cryptographic algorithms in the same way. Shor's algorithm can solve integer factorization and discrete logarithms efficiently on a sufficiently capable quantum computer [1]. It therefore threatens RSA, finite-field Diffie-Hellman, elliptic-curve Diffie-Hellman (ECDH), the Elliptic Curve Digital Signature Algorithm (ECDSA), and related systems. These mechanisms are widely used in certificates, virtual private networks, secure web sessions, device identities, code signing, and administrative access.
Grover's algorithm provides a less severe, quadratic improvement for searching symmetric-key spaces [11]. Its effect can be addressed by using sufficiently large symmetric keys and hash outputs. Consequently, PQC migration is primarily a replacement of vulnerable public-key functions rather than the encryption of all WAN traffic with a new post-quantum algorithm. AES remains suitable when an appropriate key size is used. For example, AES-256 provides a larger security margin against quantum search than AES-128. The central migration problem is how endpoints authenticate one another and establish the symmetric keys used to protect the data plane.
A CRQC capable of executing these attacks at a useful scale has not been demonstrated. Predicting its arrival requires assumptions about physical qubits, logical qubits, error correction, gate fidelity, execution time, and engineering scale. These variables make precise forecasts unreliable. A sound migration plan should therefore not depend on a single predicted Q-day. It should instead evaluate whether information could remain valuable for longer than the time available to complete migration.
This relationship can be expressed using three time intervals:

$$L + M > Q$$

where L is the required protection lifetime of the data, M is the time needed to migrate the affected system, and Q is the estimated time until a relevant quantum capability becomes available. When the sum of the data lifetime and migration time exceeds the estimated threat horizon, exposure may already exist [2]. The relationship is useful for prioritization, but it should not be treated as a precise prediction. Each term contains uncertainty and should therefore be represented by a range or reviewed under multiple scenarios.

## 2.2 Present and Future Attack Paths

The quantum threat creates two main attack paths for enterprise WANs.

The first is the loss of confidentiality through harvest-now, decrypt-later collection. An adversary records encrypted traffic and associated handshake material today. If the session keys were established solely through quantum-vulnerable public-key cryptography, the stored information might be decrypted after a CRQC becomes available. Forward secrecy protects a session when a conventional long-term private key is compromised, but it does not protect an ECDH exchange against an adversary that can later solve the underlying elliptic-curve discrete-logarithm problem. Long-term confidentiality must therefore be considered at the time the session is established.

Not every WAN flow has the same exposure. Routine operational data that loses its value quickly may present a lower harvest-now, decrypt-later risk. In contrast, research data, health information, legal records, government information, trade secrets, infrastructure designs, and long-lived identity data may require protection for many years. An enterprise should therefore evaluate the information carried by each service rather than assign the same quantum-risk rating to every encrypted tunnel.

The second attack path concerns future authentication and integrity. A CRQC could enable an attacker to derive or reproduce vulnerable signing operations, impersonate network peers, forge certificates, or defeat device identities. This risk extends beyond traffic encryption. Relevant dependencies can include:

5. IKEv2 peer-authentication certificates;
6. TLS server and client certificates;
7. administrator and device SSH keys;
8. certificate-authority hierarchies;
9. software and firmware signatures;
10. secure-boot trust anchors;
11. routing-protocol authentication dependencies; and
12. management-plane application identities.

Recorded traffic primarily creates a confidentiality concern, while signature forgery generally becomes practical only after the required quantum capability exists. Nevertheless, signature migration can take longer because public-key infrastructure, hardware trust anchors, certificate formats, validation software, and device firmware may need coordinated changes. The framework developed in this paper therefore evaluates confidentiality and authentication exposure separately.

Quantum migration can also introduce conventional security failures. Incorrect implementations may leak secret information through timing, power, cache, or fault behavior. Negotiation mechanisms may permit downgrade attacks in which two capable endpoints are induced to select a traditional-only method. Large messages may be dropped or fragmented by intermediate devices. Failed interoperability may encourage administrators to restore weaker configurations without adequate review. A quantum-ready design must address these present implementation and operational risks rather than focus only on future cryptanalysis.

## 2.3 Cryptographic Exposure in Enterprise WANs

Enterprise WAN cryptography is distributed across services, devices, protocols, trust relationships, and management components, each with its own algorithm, owner, upgrade path, and supplier dependency. A branch connection may use IKEv2 for an IPsec tunnel, certificates for gateway authentication, TLS for controller communication, Secure Shell for administration, and a separate certificate service for renewal.

For risk assessment, WAN cryptographic exposure can be divided into four planes:

**Data plane.** Protects application traffic carried through IPsec, TLS, or another encrypted overlay.

**Control plane.** Establishes tunnels, exchanges routing or reachability information, and applies network policy.
**Management plane.** Supports configuration, monitoring, logging, software distribution, and administrative access.
**Trust plane.** Provides certificates, device identities, signing keys, trust anchors, and key-management services.

This separation prevents a narrow assessment in which a post-quantum-capable tunnel is considered complete while its authentication or management dependencies remain quantum-vulnerable. It also supports more precise ownership. Network teams may control the data and control planes, while security, identity, cloud, application, and procurement teams control other dependencies.

The WAN path itself affects deployment. Branch and remote-access connections may have lower bandwidth, higher latency, greater loss, smaller effective MTUs, and less capable endpoint hardware than data-center connections. Some paths traverse network-address translation, carrier infrastructure, security inspection systems, or multiple encapsulation layers. Each added header reduces the available payload size. A post-quantum exchange that succeeds in a local laboratory may therefore fail or fragment on a production path.

## 2.4 Post-Quantum Key Establishment and Signatures

NIST FIPS 203 specifies ML-KEM, a module-lattice-based key-encapsulation mechanism [3]. Unlike a traditional Diffie-Hellman exchange, a key-encapsulation mechanism provides three principal operations:

13. Key generation produces an encapsulation key and a decapsulation key.
14. Encapsulation uses the public encapsulation key to create a ciphertext and a shared secret.
15. Decapsulation uses the private decapsulation key and ciphertext to recover the shared secret.

ML-KEM defines three parameter sets: ML-KEM-512, ML-KEM-768, and ML-KEM-1024. The parameter sets provide different security levels and message sizes. Selecting a parameter set should depend on the applicable standard, security policy, protocol profile, and required protection period. The largest parameter set is not automatically the best operational choice.

NIST also standardized ML-DSA and SLH-DSA for digital signatures [4, 5]. These algorithms address authentication and integrity rather than bulk encryption. Signature migration has different operational effects from key-establishment migration because signature and public-key sizes influence certificates, certificate chains, handshakes, storage, hardware security modules, and validation operations. This paper focuses experimentally on WAN key establishment but includes signature and certificate dependencies in the risk model. Excluding them would produce an incomplete readiness assessment.

Symmetric encryption remains responsible for protecting most application data after session establishment. PQC therefore changes how the traffic keys are established or authenticated, not how every packet is encrypted. This distinction also explains why the largest performance effects may appear during tunnel creation, certificate exchange, reauthentication, or rekeying rather than during steady-state data transfer.

## 2.5 Hybrid Migration

A hybrid key-establishment method combines a traditional secret, such as one produced by ECDH, with a post-quantum secret produced by ML-KEM. The secrets are processed through an approved combining or key-derivation method. The intended security property is that the resulting key remains protected if at least one component remains secure, subject to the assumptions and construction of the hybrid design [12].

Hybrid deployment has two practical benefits. First, it reduces dependence on a single new cryptographic assumption during the transition. Second, it can retain compatibility with established security requirements while adding protection against future quantum attacks. Hybrid designs are therefore useful during a period in which post-quantum algorithms, implementations, protocol standards, and operational experience are still developing.

Hybrid does not mean that arbitrary algorithms can safely be concatenated. The combination method, negotiation behavior, error handling, randomness, authentication, and downgrade protection all affect security. A hybrid implementation can also increase message size and processing work because it performs both traditional and post-quantum operations.

For IKEv2, RFC 9370 defines multiple key exchanges and RFC 9242 defines the IKE_INTERMEDIATE exchange used to carry additional protected material before IKE_AUTH [13, 36]. Current IETF work specifies ML-KEM alone or as an additional exchange in IKEv2 [9]. FIPS 203 gives ML-KEM ciphertext and encapsulation-key sizes from 768 to 1,568 bytes across the standardized parameter sets [3], and NIST SP 800-227 provides guidance for secure KEM use [27]. ML-KEM-768 and ML-KEM-1024 messages can exceed restricted path MTUs after protocol and encapsulation overhead are added. Traditional-only negotiation also remains a downgrade concern when peers support both traditional and post-quantum configurations.

TLS 1.3 is undergoing a similar transition. Current IETF work defines hybrid groups that combine ECDHE with ML-KEM [8]. These mechanisms can protect TLS key establishment against a future quantum attacker while preserving a traditional component. However, larger client and server handshake messages can affect connection establishment, especially on lossy or high-delay paths. Testing must therefore cover end-to-end protocol behavior rather than only the execution time of the ML-KEM operations.

### 2.6 Standards and Government Migration Direction

The publication of FIPS 203, FIPS 204, and FIPS 205 changed PQC migration from pre-standard experimentation to standards-based implementation [3-5]. Algorithm standards do not finalize every protocol binding or establish that a particular implementation, module, or deployment is suitable. Organizations must distinguish finalized algorithms, approved protocol profiles, active Internet-Drafts, experimental implementations, and validated production modules.

NIST IR 8547 identifies quantum-vulnerable standards and proposes a transition toward quantum-resistant key-establishment and signature mechanisms [14]. The NIST publication record still listed IR 8547 as an Initial Public Draft on August 9, 2026. It can inform planning, but its dates and transition details should be rechecked immediately before submission.

U.S. federal policy emphasizes inventory, accountable ownership, and risk-based sequencing. OMB M-23-02 requires inventories of active cryptographic systems, including protection lifetime and system impact [6]. Executive Order 14412 requires PQC key establishment for federal high-value assets and high-impact systems by December 31, 2030, and PQC digital signatures by December 31, 2031 [37]. OMB M-26-15 requires a migration plan within 120 days, prioritized key-establishment migration through 2030, signature migration in 2031, and risk-based completion planning toward 2035 [7]. Private enterprises can use these milestones as governance benchmarks without treating them as universally binding law.

Other governments have adopted comparable staged approaches. The United Kingdom's National Cyber Security Centre recommends completing discovery and initial planning by 2028, migrating the highest-priority services by 2031, and completing migration by 2035 [15]. The guidance also recognizes supplier dependencies, long-lived infrastructure, coexistence between traditional and post-quantum mechanisms, and the need for rollback and business-continuity planning. These concerns closely match the operational conditions of enterprise WANs.

Across these directives, five common activities emerge:

16. discover cryptographic use;
17. identify sensitive and long-lived information;
18. map technical and supplier dependencies;
19. prioritize migration according to risk; and
20. validate new cryptography before broad deployment.

Together, these directives establish a common program of discovery, data and system prioritization, dependency mapping, accountable planning, and validation. The unresolved WAN-level questions are how to compare individual services, how to set service-specific acceptance gates, and how to verify that hybrid negotiation cannot silently downgrade.

### 2.7 Scope of "Quantum-Ready"

In this paper, quantum-ready is a post-deployment service state. The relevant cryptographic dependencies are known; confidentiality and authentication exposure have been assessed; required post-quantum or hybrid controls are enforced; prohibited fallback is controlled; interoperability and path behavior have been tested; operational gates are satisfied; monitoring and rollback exist; and remaining exposure or exceptions are documented. This differs from an MRS deployment-prepared rating, which describes the ability to begin or expand migration.
A WAN service is therefore considered quantum-ready only when:

21. its relevant cryptographic dependencies are known;
22. its confidentiality and authentication exposure has been assessed;
23. its selected protection follows applicable standards;
24. traditional-only fallback is controlled;
25. interoperability and path behavior have been tested;
26. operational thresholds have been satisfied;
27. monitoring and rollback procedures exist; and
28. residual exposure, accepted risk, and remaining dependencies are documented.

Algorithm availability is necessary, but a service is quantum-ready only when configuration, behavior, evidence, and remaining dependencies support that conclusion.

## 3 Related Work and Research Gap

Research relevant to quantum-ready WAN migration spans organizational transition, cryptographic discovery, cryptographic agility, protocol design, measurement, and risk communication. The framework adopts established work in each stream and focuses on the unresolved step of joining those inputs at the WAN-service level.

### 3.1 Organizational Migration and Cryptographic Agility

Joseph et al. frame post-quantum migration as a multi-year organizational problem involving inventories, standards, suppliers, applications, and governance [2]. Ott et al. identify related crypto-agility research challenges [10], while Hasan et al. use dependency analysis to expose relationships that a flat inventory misses [17]. NIST's crypto-agility guidance extends the scope across protocols, software, hardware, firmware, and infrastructure [19]. This paper adopts that dependency-oriented view and narrows the assessment unit to an operational WAN service. Their main limitation for the present problem is granularity: none defines how a WAN operator should keep exposure priority, deployment preparation, and evidence quality separate when comparing individual tunnels or application relationships.

### 3.2 Cryptographic Discovery and Dependency Mapping

The NIST NCCoE migration project examines cryptographic discovery and inventory construction [18]. Configuration analysis, packet inspection, software inventories, and certificate records can show where quantum-vulnerable cryptography is used. The framework uses these outputs as evidence, then adds service ownership, timing, impact, path, dependency, and validation fields needed for sequencing.

### 3.3 Performance of Post-Quantum TLS

Prior TLS studies show that post-quantum performance depends on algorithm size, certificate chains, network conditions, and endpoint capacity. Sikeridis et al. study post-quantum authentication in TLS 1.3 [20]; Sosnowski et al. examine constrained paths [21]; and Montenegro et al. provide a reproducible comparison framework for classical, hybrid, and post-quantum TLS [25]. The planned experiment does not claim novelty from another handshake benchmark alone. Its distinguishing measurements are branch-class versus server-class capacity, rekey and establishment-rate limits, and negative testing tied to service acceptance gates.

### 3.4 Performance of Post-Quantum IPsec and IKEv2

Bae et al. measure post-quantum IKEv2 execution time and packet size [22]. Mutlugun et al. evaluate satellite-network conditions [23], and Twardokus et al. examine constrained-network behavior [24]. Their results motivate the loss, MTU, capacity, and fragmentation conditions in Section 7. This paper adds an enterprise decision layer: services are selected for measurement by exposure priority, and the measurements feed predefined security and operational gates.

### 3.5 Hybrid Key Establishment and Protocol Transition

ETSI describes hybrid key exchange as a transition technique [12]. RFC 9370 defines multiple IKEv2 key exchanges [13], RFC 9242 defines IKE_INTERMEDIATE [36], and current IETF work specifies ML-KEM profiles for TLS 1.3 and IKEv2 [8, 9]. Bhargavan et al. show why downgrade resilience must be treated as a protocol property rather than inferred from successful negotiation [26]. The migration framework adopts that principle by making prohibited-fallback tests a non-compensable gate.

### 3.6 Policy and Operational Guidance

Government guidance provides migration outcomes and milestones. OMB requires inventories, accountable plans, and risk-based sequencing for U.S. federal systems [6, 7, 37], while the UK NCSC provides staged targets through 2035 [15]. The framework translates those program requirements into service records, waves, and evidence gates without treating policy milestones as a prediction of Q-day.

### 3.7 Synthesis

Table 1 summarizes the literature streams and the remaining WAN decision.

**Table 1. Related research and the remaining enterprise WAN question**

| Research stream | Main contribution | Remaining WAN question |
|---|---|---|
| Migration frameworks | Organizational steps and dependencies | Which WAN service should migrate first? |
| Cryptographic discovery | Vulnerable algorithms and assets | How should inventory become exposure-priority bands? |
| Crypto-agility | Safe algorithm replacement | How should urgency and readiness be separated? |
| TLS performance | Handshake and certificate costs | Is the profile acceptable for a specific service? |
| IPsec performance | Path and IKEv2 effects | When should a tunnel receive deployment approval? |
| Hybrid specifications | Combined key-establishment mechanisms | How are fallback and downgrade controlled? |
| Government guidance | Outcomes and timelines | How are objectives implemented at WAN service level? |

### 3.8 Research Gap

The literature leaves an integration gap across six functions: WAN-specific threat modeling, service prioritization, TLS and IKEv2 validation under representative paths, business-defined acceptance gates, staged migration action, and documented residual exposure. The composite QES introduced below is intentionally limited to ordinal ranking and banding. It is not a probability or an interval measure of risk; critiques of unjustified security-scoring formulas make that boundary important [29, 38].

## 4 Enterprise WAN Threat Model

The threat model defines the assets, adversaries, trust boundaries, and failure events considered by the framework. It covers present collection attacks, future quantum-enabled attacks, and conventional failures introduced during migration.

### 4.1 System Model

A protected business flow passes from an application through Gateway A, across an untrusted carrier, Internet, wireless, or cloud path, and through Gateway B to a data-center, cloud, partner, or branch service. Management and public-key infrastructure (PKI) services configure and authenticate the endpoints. The adversary may record, delay, drop, replay, reorder, or alter traffic on the untrusted path. Figure 1 shows the principal trust relationships.

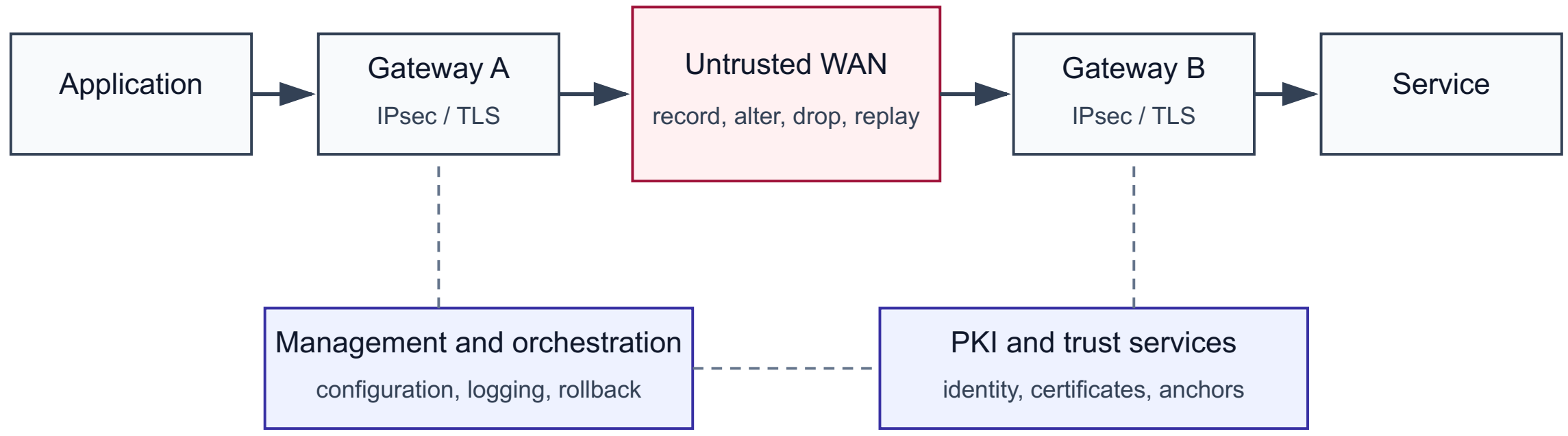

**Fig. 1. Enterprise WAN system and principal trust relationships.**

### 4.2 Protected Assets

29. Long-lived business, research, personal, regulated, and backup data.
30. Session keys, key-establishment material, and rekey state.
31. Gateway, server, user, and administrative identities.
32. Network configuration, routing policy, and security policy.
33. Management, orchestration, certificate, and trust infrastructure.
34. Availability of business-critical connectivity and control services.

### 4.3 Security Objectives

Table 2 defines the security objectives used by the threat model.

**Table 2. Security objectives**

| Objective | Required property |
|---|---|
| Long-term confidentiality | Recorded traffic remains protected for its required lifetime. |
| Peer authentication | Endpoints accept only authorized peers and identities. |
| Integrity | Unauthorized modification, injection, and replay are detected. |
| Downgrade resistance | A required hybrid or post-quantum mode cannot silently fall back. |
| Availability | Migration does not create unacceptable connection or rekey failures. |
| Cryptographic agility | Algorithms and profiles can be replaced without uncontrolled disruption. |

### 4.4 Adversary Classes

A1 - Present passive collector. Records encrypted WAN traffic and associated handshakes for later analysis.
A2 - Present active network attacker. Alters, drops, replays, or delays traffic and attempts negotiation or implementation attacks.
A3 - Future quantum-capable attacker. Uses a CRQC to attack vulnerable key establishment or signature mechanisms.

A4 - Migration-aware attacker. Exploits inconsistent deployment, optional fallback, configuration drift, or weak rollback procedures.

### 4.5 Trust Boundaries

35. TB1: Endpoint boundary between applications and the gateway or TLS stack.
36. TB2: Underlay boundary between enterprise-controlled equipment and the external path.
37. TB3: Cloud or third-party boundary where configuration and evidence may be supplier controlled.
38. TB4: Management boundary connecting controllers, administrators, logging, and software distribution.
39. TB5: Trust boundary containing certificate authorities, enrollment, trust anchors, and signing services.

### 4.6 Threat Events

Table 3 links each threat event to the evidence required for assessment.

**Table 3. Threat events and evidence**

| ID | Threat event | Affected objective | Required evidence |
|---|---|---|---|
| T1 | Harvest-now, decrypt-later collection | Confidentiality | Data lifetime, path exposure, and enforced key establishment |
| T2 | Quantum-enabled peer impersonation | Authentication, integrity | Certificate, signature, and trust-dependency inventory |
| T3 | Downgrade to traditional-only negotiation | Confidentiality, authentication | Negative tests, logs, and fallback policy |
| T4 | Incomplete dependency migration | All objectives | Dependency graph and configuration verification |
| T5 | Fragmentation or handshake failure | Availability | Packet captures, MTU tests, retransmission and failure rates |
| T6 | Resource exhaustion during setup or rekey | Availability | CPU, memory, capacity, and recovery tests |
| T7 | Migration drift or insecure rollback | All objectives | Change records, monitoring, exceptions, and rollback tests |

### 4.7 Migration-Induced Risk

Migration may create outages, certificate failures, packet-size problems, inconsistent configurations, uncontrolled fallback, weak random-number use, incorrect KEM integration, or monitoring gaps. Reliance on an Internet-Draft may also create identifier or behavior changes between implementation versions. These conventional risks are assessed alongside the future quantum threat rather than treated as reasons to postpone planning.

### 4.8 Assumptions

40. No public CRQC capable of breaking deployed RSA or elliptic-curve cryptography is assumed to exist at present.
41. Adversaries can record traffic traversing external or shared WAN paths.

42. Standardized PQC algorithms are used according to their stated security assumptions and implementation guidance.
43. Appropriately sized symmetric cryptography remains usable, subject to current policy and implementation quality.
44. Traditional and post-quantum mechanisms will coexist during a multi-year transition.
45. WAN services have distributed technical owners, suppliers, and maintenance constraints.
46. Operational failure may cause personnel to restore a weaker configuration unless rollback is controlled.

### 4.9 Exclusions

The paper does not propose a new post-quantum algorithm, provide formal protocol proofs, evaluate quantum key distribution, or model physical device compromise, general malware, or traffic-analysis attacks unrelated to cryptographic transition. Opaque proprietary SD-WAN internals are assessed only through observable configuration and behavior. Full post-quantum certificate-chain performance is outside the core key-establishment experiment.

### 4.10 Threat-to-Measurement Traceability

Table 4 connects the threat model to assessment fields and validation evidence.

**Table 4. Threat-to-measurement traceability**

| Threat | Assessment and evidence fields | Validation measurements |
|---|---|---|
| T1 | L_C, M_C, T_C, E_C, V_K | Negotiated profile, traffic capture, and fallback rejection |
| T2 | L_A, M_A, T_A, I_A, V_A, D | Certificate, signature, trust inventory, and authentication tests |
| T3 | V_K or V_A; control-policy evidence | Negative negotiation tests, alerts, and downgrade evidence [26] |
| T4 | D; inventory and MRS evidence | Dependency coverage and configuration review |
| T5 | E_C or E_A; operational-validation evidence | Handshake size, MTU, fragments, retransmissions, and failure |
| T6 | I_C or I_A; capacity evidence | CPU, memory, establishment capacity, and recovery |
| T7 | Governance MRS dimension and ECS | Drift detection, exception expiry, and rollback evidence |

## 5 Quantum-Exposure Prioritization Model

The assessment model produces four separate outputs: confidentiality and authentication Quantum Exposure Scores (QES_C and QES_A), a Migration Readiness Score (MRS), and an Evidence Confidence Score (ECS). QES is an ordinal prioritization index for comparable services assessed under the same rubric. It is not a probability of compromise, a monetary loss estimate, or an interval-scale measure of risk. MRS and ECS remain separate so that supplier readiness or weak evidence cannot lower the underlying exposure priority.

### 5.1 Assessment Unit

The assessment unit is a WAN service: a business flow or application relationship, its tunnel or TLS session, both endpoints, and its management, identity, trust, software, hardware, and supplier dependencies. Services may be grouped only when these properties and their business effects are materially identical.
Table 5 defines the symbols, ranges, and decision meanings used below.

**Table 5. Notation used in the assessment and experiment**

| Symbol | Range | Meaning |
|---|---|---|
| L_C, L_A | years, >= 0 | Required remaining protection lifetime for confidentiality and authentication |
| M_C, M_A | years, >= 0 | Lead time to deploy and verify key-establishment or authentication migration |
| Q | years, > 0 | Planning horizon to a CRQC; scenario input, not a forecast |
| H_C, H_A | years | Timing margin L + M - Q |
| T_C, T_A | 0-5 | Smoothed timing urgency derived from H and u |
| I_C, I_A | 0-5 ordinal | Confidentiality and authentication impact |
| V_K, V_A | 0-5 ordinal | Key-establishment and authentication control-state vulnerability |
| E_C, E_A | 0-5 ordinal | Passive-collection and active-authentication path exposure |
| D | 0-5 ordinal | Shared dependency concentration |
| QES_C, QES_A | 0-100 ordinal index | Exposure-priority scores used for ranking and banding |
| MRS | 0-100 in steps of 4 | Migration preparation across five maturity dimensions |
| ECS | 0-100 | Weighted confidence in the evidence supporting the assessment |
| p*, n | probability; count | Service failure limit and number of establishment attempts |

### 5.2 Timing Urgency

For confidentiality, L_C is the remaining data-protection lifetime, M_C is the estimated time to deploy and verify the required key-establishment control, and Q is the assumed number of years until a CRQC. The timing margin is H_C = L_C + M_C - Q. For authentication, L_A is the required lifetime of the identity, trust anchor, signed artifact, or service, and M_A is the time to migrate its signature, certificate, validation, hardware, and trust dependencies. The authentication margin is H_A = L_A + M_A - Q.
For dimension x in {C, A}, T_x = 5 x clip((H_x + u) / (2u), 0, 1), with u > 0 and u = 5 years by default. The values Q = 5, 10, and 15 years are planning scenarios rather than predictions.
With u = 5, a margin of -5 years maps to 0, the Mosca boundary H_x = 0 maps to 2.5, and a margin of +5 years maps to 5. Mapping the boundary to the midpoint is a conservative smoothing choice: services near either side of H_x = 0 remain visible for review. Saturation also means that the scenarios may not change the band for very long-lived services, so the report shows both scores and band stability.

### 5.3 Exposure Dimensions

Timing urgency (T), business impact (I), cryptographic control-state vulnerability (V), path exposure (E), and dependency concentration (D) use ordered levels from 0 to 5. Except for T, these levels are ordinal judgements. Their spacing is not assumed to represent equal amounts of risk.
Impact: 0 is negligible; 1 limited; 2 material but localized; 3 serious operational, contractual, financial, or regulatory effect; 4 severe enterprise or critical-service effect; and 5 catastrophic, systemic, safety, or national effect. The six levels extend the low, moderate, and high impact concept in FIPS 199 [30] to support enterprise sequencing; they do not replace a formal FIPS 199 categorization. $I_C$ and $I_A$ may differ because loss of confidentiality and loss of authentication can have different consequences.
Key-establishment control state ($V_K$): 0 means the required post-quantum or hybrid profile is enforced, independently verified, and rejects prohibited fallback; 1 has complete enforcement with only minor production-evidence gaps; 2 uses controlled and monitored fallback under an expiring exception; 3 is partial or inconsistent deployment; 4 relies on an unverified compensating mechanism; and 5 uses traditional quantum-vulnerable public-key key establishment only.
Authentication control state ($V_A$): 0 means the required post-quantum signature and trust path is enforced and verified end to end; 1 has complete enforcement with minor evidence gaps; 2 uses a controlled dual-certificate or fallback state; 3 has a partially migrated chain, peer set, renewal path, or validation stack; 4 relies on an unverified compensating mechanism or supplier statement; and 5 uses only quantum-vulnerable signatures or trust anchors.
Path exposure (E): 0 represents a controlled enclave; 1 a tightly controlled private path; 2 a managed private service; 3 a shared carrier or cloud path; 4 a public, wireless, or partner path; and 5 a highly exposed or targeted path. $E_C$ and $E_A$ use the same rubric but may differ when passive collection and active authentication attack surfaces differ.
Dependency concentration (D): 0 indicates no shared dependency; 1 one or two local dependencies; 2 several dependencies within a small group; 3 multi-site dependencies; 4 many critical dependencies; and 5 an enterprise-wide controller, trust, identity, or gateway dependency. D is shared across $QES_C$ and $QES_A$ unless the service is split into separate assessment records.

### 5.4 Exposure Equations

$QES_C = 6T_C + 5I_C + 4V_K + 2E_C + 3D$.
$QES_A = 5T_A + 6I_A + 4V_A + 2E_A + 3D$, and $QES = \max(QES_C, QES_A)$.
The integer coefficients are proposed decision defaults and make the contribution of each factor visible. Because I, V, E, and D are ordinal, QES supports ordering and band assignment only. Point differences are not interpreted as quantities of risk removed. The planned validation varies each coefficient by +/-20%, perturbs each ordinal input by +/-1 level, varies band cut points by +/-5 points, and reports rank or band instability [38].

### 5.5 Priority Bands and Scenarios

Table 6 defines the ordinal exposure-priority bands.

**Table 6. Quantum Exposure Score priority bands**

| QES | Priority band | Interpretation |
|---|---|---|
| 80-100 | Critical | Immediate treatment and controlled migration action |
| 60-79 | High | Accelerated migration and testing |

| QES | Priority band | Interpretation |
|---|---|---|
| 40-59 | Moderate | Planned migration and readiness improvement |
| 0-39 | Low | Monitor and address through lifecycle planning |

Each service is reported as QES_5, QES_10, and QES_15. The primary output is the ordering and band under each scenario; a numerical difference between two scores is not treated as a calibrated distance. Section 9 reports cases that are stable and cases that cross a band when Q changes.

### 5.6 Migration Readiness Score

MRS measures five dimensions: inventory completeness, standards and implementation support, interoperability evidence, operational validation, and governance and observability. Equal weighting is a transparent default to be tested with practitioners. Each dimension uses a maturity scale: 0 unknown; 1 informal or unverified; 2 documented plan or supplier statement; 3 partial laboratory evidence; 4 controlled pilot evidence; and 5 monitored production evidence.
Let R be the sum of the five maturity levels. MRS = 4R, where R is an integer from 0 to 25. Table 7 places attainable MRS values into preparation states.

**Table 7. Migration Readiness Score preparation states**

| MRS | Preparation state |
|---|---|
| 76-100 (R = 19-25) | Deployment-prepared |
| 52-72 (R = 13-18) | Partial |
| 0-48 (R = 0-12) | Low |

### 5.7 Exposure-Readiness Decision Matrix

Table 8 maps urgency and deployment preparation to action.

**Table 8. Exposure-priority and readiness actions**

| Exposure priority | Deployment-prepared | Partial readiness | Low readiness |
|---|---|---|---|
| Critical | Migrate immediately through approved gates | Urgent pilot and temporary protection | Executive escalation, containment, and capability program |
| High | Accelerated deployment | Priority testing | Supplier escalation, compensating controls, and funded remediation |
| Moderate | Planned deployment or quick win | Scheduled pilot | Inventory and readiness improvement |
| Low | Lifecycle migration | Monitor and prepare | Monitor and document |

The constructs mark different decisions. A QES priority band expresses urgency; an MRS state expresses deployment preparation; a wave orders services; and a phase records progress through the migration workflow. Wave 0 denotes immediate treatment, Wave 1 accelerated migration, Wave 2 planned migration, Wave 3 lifecycle migration, and Wave 4 monitoring. Frequent rekey of quantum-vulnerable ECDH does not prevent later decryption of recorded handshakes.

### 5.8 Evidence Confidence Score

Evidence is scored from 0 to 5: 0 unknown; 1 assumption; 2 owner or supplier statement; 3 current inventory or configuration; 4 automated discovery or laboratory measurement; and 5 direct independent evidence or monitored production evidence. Eight evidence items are graded: protection lifetime (weight 0.15), migration lead time (0.10), impact (0.15), cryptographic control state (0.15), path exposure (0.10), dependency concentration (0.10), readiness dimensions (0.15), and protocol or configuration verification (0.10). With evidence grades e_j, ECS = 20 x sum(w_j e_j). Scores of 80 or more are High, 60-79 Moderate, and below 60 Low. Unknown evidence creates a conservative provisional assessment and cannot justify deferral.

### 5.9 Deployment Gates

Security gates require an approved profile, rejection of prohibited fallback, downgrade alerts, reviewed authentication dependencies, no unapproved bypass, correct randomness and key derivation, and packet-level confirmation of the negotiated state.
Table 9 states the operational gates that must be set before testing.

**Table 9. Operational deployment gates**

| Measure | Gate |
|---|---|
| Latency | p50, p95, and p99 remain within the service-level objective (SLO) |
| Failure | Upper confidence bound remains within the error budget |
| Packet handling | No uncontrolled fragmentation; retransmission remains acceptable |
| Rekey | Reliable completion without unacceptable interruption |
| Processor | Required CPU headroom remains available |
| Memory | Peak and steady memory remain within capacity |
| Throughput | Application throughput remains within the service SLO |
| Rollback | Returns to the last approved state without enabling prohibited fallback |

Governance gates require an accountable owner, current inventory record, logs and alerts, an expiring exception process, tested rollback, continuing monitoring, and a reassessment date. Security, operational, and governance gates must all pass.

### 5.10 Residual Exposure and Reassessment

After a control is deployed, closure is determined by evidence and gates rather than by subtracting ordinal QES points. The residual-exposure state is Open when traditional-only or unverified protection remains; Transitional when the required control is deployed but fallback, evidence, path, or dependency gaps remain; Prospectively closed when new sessions enforce and verify the required quantum-resistant protection with no prohibited fallback; and Quantum-ready when both confidentiality and authentication objectives satisfy all gates and remaining legacy exposure or exceptions are documented. Previously recorded traffic may remain exposed even after prospective confidentiality is closed.

### 5.11 Assessment Procedure

1. Define the WAN service and business owner.
2. Record data and authentication protection lifetimes.
3. Map endpoints, paths, algorithms, identities, trust, management, and supplier dependencies.
4. Estimate migration lead time and select Q scenarios.
5. Score confidentiality and authentication exposure.
6. Score migration readiness and evidence confidence.
7. Assign the provisional migration wave.
8. Select a target control and predefined acceptance gates.
9. Execute laboratory and pilot validation.
10. Deploy, record the residual-exposure state, and schedule reassessment.

### 5.12 Illustrative Calculation

A research-data tunnel has $L_C = 15$ years, $M_C = 3$ years, and $Q = 10$ years. Thus $H_C = 8$ and $T_C = 5$. With $I_C = 4$, $V_K = 5$, $E_C = 4$, and $D = 3$, $QES_C = 87$, which is in the Critical priority band. Readiness inputs of 3, 2, 1, 1, and 2 produce MRS = 36, or Low. The decision is Wave 0 treatment, executive and supplier escalation, and an urgent controlled pilot. The score orders work; it is not an estimate of compromise probability.

## 6 Enterprise WAN Migration Framework

The framework converts the assessment into a controlled migration program. Each service passes through discovery, assessment, architecture, laboratory validation, pilot, deployment, retirement, and reassessment. Evidence determines progression, while a failed gate returns the service to remediation as shown in Figure 2.

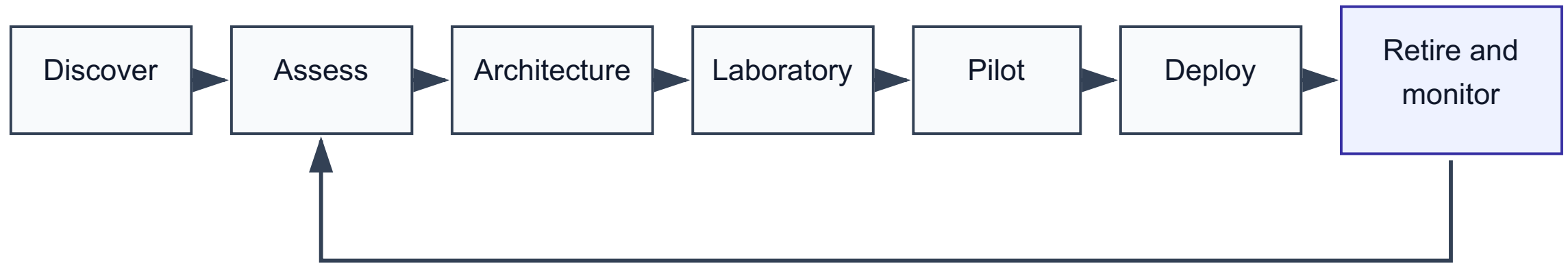

**Fig. 2. Enterprise WAN post-quantum migration lifecycle.**

### 6.1 Migration Principles

1. Assess services, not isolated devices. The gateway, applications, peer, identities, path, management system, and business owner form one assessment unit.
2. Keep exposure and readiness separate. Low product availability creates an urgent engineering problem; it does not lower quantum exposure.

3. Separate confidentiality from authentication. Hybrid key establishment may protect recorded traffic while certificates and trust services remain vulnerable.
4. Prefer standardized algorithms and defined protocol profiles. Algorithm standardization does not by itself establish protocol interoperability or policy approval.
5. Prevent silent fallback. A required hybrid or post-quantum state must be enforced and observable.
6. Validate operational behavior before broad deployment, including size, fragmentation, retransmission, resources, rekey, failure, and monitoring.
7. Make rollback controlled and auditable. Rollback restores the last approved state rather than permanently enabling a weaker mode.

## 6.2 Migration Phases and Decision Gates

Table 10 defines the workflow evidence and exit condition for each phase.

**Table 10. Migration phases, evidence, and exit conditions**

| Phase | Main activity | Required evidence | Exit condition |
|---|---|---|---|
| 0 Governance | Define scope, ownership, approved profiles, scenarios, exceptions, and dates | Program charter and named owners | CISO and business owner approve scope |
| 1 Discovery | Identify algorithms, tunnels, certificates, libraries, hardware, management, and dependencies | Configuration, scans, captures, and owner confirmation | Critical services have owners and documented cryptographic paths |
| 2 Assessment | Calculate QES, MRS, and ECS; assign a priority band and wave | Scoring record and supporting evidence | Risk and technical owners confirm classification |
| 3 Architecture | Select hybrid, post-quantum, or temporary controls; define fallback, MTU, logging, and rollback | Architecture decision record | Security architecture approves a testable design |
| 4 Laboratory | Test interoperability, performance, failures, downgrade, and rekey | Reproducible test report | Security and operational gates pass |
| 5 Pilot | Deploy to a bounded representative production group | Production telemetry and incident record | Pilot remains within predefined limits |
| 6 Deployment | Expand through predefined service or site cohorts | Change records, monitoring, and exception status | Target population is migrated |
| 7 Retirement | Disable traditional-only modes, remove obsolete configuration, and monitor residual exposure | Configuration verification and negative tests | Legacy paths are rejected or time-limited; residual-exposure state is recorded |

## 6.3 Selecting a Migration Control

Table 11 compares the available migration control patterns.

**Table 11. Migration control patterns**

| Control pattern | Appropriate use | Main limitation |
|---|---|---|
| Traditional with monitoring | Low-risk service awaiting a supported path | Does not address quantum exposure |
| Compensating symmetric control | Temporary protection for selected IKEv2 environments | Key distribution and lifecycle burden |
| Hybrid key establishment | Transition where traditional and post-quantum components can be combined | Requires correct composition, interoperability, and downgrade protection |

| Control pattern | Appropriate use | Main limitation |
|---|---|---|
| Post-quantum-only profile | Target state when standards, policy, peer support, and evidence are mature | Not yet practical for every protocol or dependency |

Hybrid key establishment is intended to retain protection when at least one independent component remains secure, provided the secrets are combined correctly and negotiation is authenticated [8, 9, 12, 13]. For selected IKEv2 services, a high-entropy post-quantum preshared key can provide an interim layer of protection for derived traffic keys [31, 32]. This mechanism creates key-distribution and rotation burdens and does not remove the need to migrate authentication.

Control selection considers the protected objective, profile maturity, peer interoperability, path tolerance, authentication dependencies, enforceable fallback, observability, and safe rollback.

### 6.4 Constructing Migration Waves

1. Apply the exposure-readiness matrix to assign the initial wave.
2. Migrate shared trust, management, and identity dependencies before dependent services where sequencing requires it.
3. Within a tier, sort by the highest QES under the conservative scenario and by migration lead time.
4. Treat low-confidence assessments as investigation priorities, not evidence for deferral.
5. Group deployment only when configuration, path, business effect, and rollback conditions are sufficiently similar.
6. Keep Critical low-readiness services in Wave 0 while containment, supplier escalation, and engineering proceed.

### 6.5 Pilot and Progressive Deployment

1. Record the traditional baseline using the measurements in Section 7.
2. Enable the target profile on a bounded set of representative peers.
3. Verify negotiated algorithms using independent configuration, logs, or packet captures.
4. Test rejection of prohibited fallback and malformed or incomplete negotiation.
5. Observe latency, retransmission, fragmentation, CPU, memory, throughput, rekey, and failure.
6. Exercise restart, certificate renewal, peer unavailability, and rollback.
7. Expand only after all predefined gates remain satisfied during the observation period.

Cohort size and observation duration are declared before the pilot so that success is not defined after results are known.

### 6.6 Rollback and Exception Management

Rollback is an authorized change to a documented and tested state. Fallback is a protocol negotiation that selects a weaker mode. Rollback may be necessary for service continuity, but silent fallback is prohibited when policy requires post-quantum protection.

47. Every exception identifies the service, owner, QES, MRS, and ECS.
48. It records why migration cannot proceed and which temporary controls apply.

49. It defines the permitted cryptographic state, monitoring, supplier commitment, expiration, and reassessment date.
50. An expired exception is a control failure and requires a new risk decision before extension.

### 6.7 Governance and Responsibility

Table 12 assigns primary ownership across the program.

**Table 12. Migration roles**

| Role | Primary responsibility |
|---|---|
| Executive risk owner | Approves priorities, funding, and residual-risk acceptance |
| Chief information security officer (CISO) or program owner | Maintains the framework, profiles, and migration dashboard |
| WAN engineering | Discovers tunnels, implements configurations, and conducts network tests |
| PKI and identity | Migrates certificates, trust anchors, signing dependencies, and enrollment |
| Application and data owner | Defines protection lifetime, business impact, and service limits |
| Security operations | Monitors negotiation, downgrade attempts, failures, and drift |
| Procurement | Translates requirements into contracts and supplier roadmaps |
| Audit or assurance | Reviews evidence, exceptions, and claimed completion |

### 6.8 Procurement Requirements

Requests for proposal and contract renewals should require suppliers to state:

51. supported algorithms and exact protocol profiles;
52. availability and enforcement of hybrid and post-quantum-only modes;
53. whether traditional-only fallback can be disabled;
54. applicable validation or conformance status;
55. maximum handshake and certificate-chain sizes;
56. measured processor, memory, latency, throughput, MTU, fragmentation, retransmission, and rekey behavior;
57. interoperability evidence with independent implementations;
58. logging and alerts for negotiated algorithms and downgrade events;
59. software, firmware, hardware, and end-of-support dependencies; and
60. exportable configuration and cryptographic inventory information.

A supplier statement is readiness-planning evidence, not proof that a particular WAN service will operate correctly.

### 6.9 Alignment with Standards and Government Guidance

Table 13 maps external direction to the framework's service-level controls.

Table 13. External direction and framework response

| External direction | Framework response |
| --- | --- |
| NIST FIPS 203-205 [3-5] | Uses standardized post-quantum algorithms as the cryptographic basis |
| NIST NCCoE migration work [18] | Requires discovery, dependency mapping, interoperability, and performance evidence |
| NIST crypto-agility guidance [19] | Includes replaceable profiles, observability, controlled transition, and retirement |
| EO 14412 and OMB M-26-15 [7, 37] | Maps accountable planning to key-establishment migration by 2030, signature migration by 2031, and risk-based completion toward 2035 |
| UK NCSC timeline [15] | Supports discovery and planning by 2028, priority migration by 2031, and completion planning toward 2035 |

The applicable dates and duties depend on jurisdiction and sector. The mapping is an implementation aid, not a claim that one directive applies globally or that use of the framework establishes compliance.

### 6.10 Migration Evidence Package

61. Service and dependency description.
62. Cryptographic inventory and QES, MRS, and ECS calculations.
63. Architecture and control-selection decision.
64. Laboratory and production-pilot results.
65. Fallback, downgrade, deployment, and rollback evidence.
66. Open exceptions and post-deployment residual-risk assessment.

## 7 Experimental Design and Methodology

The complete evaluation design has two parts. Controlled experiments would measure the effect of hybrid ML-KEM key establishment on TLS 1.3 and IKEv2/IPsec, while a framework evaluation would examine practitioner consistency and sensitivity to model assumptions. The retained evidence described in Section 8 is only a partial aggregate TLS summary and is not represented as execution of this complete design.

### 7.1 Evaluation Questions and Hypotheses

67. EQ1: What latency, traffic, processor, and memory overhead is introduced by hybrid ML-KEM key establishment?
68. EQ2: How do delay, loss, bandwidth, and MTU affect hybrid TLS and IKEv2 establishment?
69. EQ3: Does the hybrid profile increase fragmentation, retransmission, timeout, or rekey failure?
70. EQ4: Can implementations enforce the required profile and reject prohibited traditional-only fallback?
71. EQ5: Do the measurements change MRS or the migration-wave decision?
72. EQ6: Can independent practitioners apply the prioritization model consistently?

H1: Hybrid key establishment increases handshake size and computational cost relative to the traditional baseline.

H2: Its operational effect increases under packet loss and constrained MTU conditions.

H3: The same profile may be acceptable for one WAN service and unacceptable for another because hardware, path conditions, and service objectives differ.

### 7.2 Testbed Architecture

The intended complete testbed places two protocol endpoints on opposite sides of a controlled WAN impairment system, while an independent observation host collects captures, system metrics, and protocol logs. Figure 3 shows this intended structure.

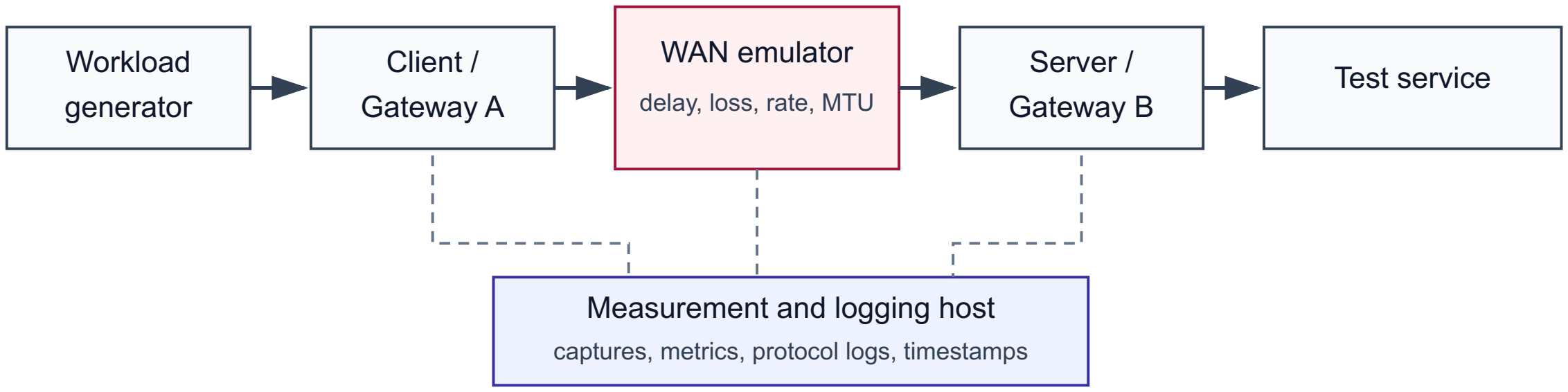

Fig. 3. Controlled WAN testbed.

The emulator in the full design applies symmetric delay, rate limits, packet loss, and MTU constraints on an isolated network. A complete empirical artifact must report hardware, operating system, kernel, firmware, cryptographic acceleration, interface offloads, compiler settings, virtualization configuration, and achieved impairment distributions. Those materials are not part of the retained aggregate artifact.

### 7.3 Protocol Profiles

Table 14 defines the classical and hybrid protocol profiles.

Table 14. Experimental protocol profiles

| Profile | Protocol | Key establishment | Purpose |
|---|---|---|---|
| TLS-C | TLS 1.3 | X25519 | Traditional baseline |
| TLS-H | TLS 1.3 | X25519MLKEM768 | Hybrid comparison |
| IKE-C | IKEv2/IPsec | Curve25519 | Traditional baseline |
| IKE-H | IKEv2/IPsec | Curve25519 and ML-KEM-768 | Hybrid comparison |

ML-KEM-768 is standardized in FIPS 203 and appears in current TLS and IKEv2 work [3, 8, 9]. TLS 1.3, IKEv2, and Curve25519 baselines follow RFCs 8446, 7296, and 8031 [33-35]. In the complete design, symmetric ciphers, certificate chains, authentication, traffic selectors, payloads, and rekey policy remain constant within each comparison; TLS resumption and zero round trip time are disabled.

The TLS and IKEv2 ML-KEM bindings cited in [8] and [9] remain Internet-Drafts and works in progress as of August 10, 2026. A full experiment must record the exact revision, identifiers, implementation commit, and collection date. Traditional certificate authentication remains constant in the proposed comparison, so the design does not claim complete post-quantum authentication.

### 7.4 Implementation Selection

Before a full execution, a preregistered record must name every protocol implementation, version, commit, cryptographic provider, build option, operating system, and hardware target. At least two independent implementations should be used where the target profiles interoperate. The limited artifact records author-reported component versions but does not contain the build evidence or raw observations needed to verify protocol execution.

### 7.5 WAN Conditions

Table 15 defines the principal path and queue conditions.

**Table 15. Controlled WAN factors**

| Factor | Values |
|---|---|
| Round-trip time | Approximately 0, 20, 80, and 160 ms |
| Packet loss (main matrix) | 0, 0.1%, 0.5%, and 1.0% per direction |
| MTU | 1280, 1400, and 1500 bytes |
| Bandwidth | 10, 100, and 1000 Mbit/s |
| Loss process | Independent Bernoulli in the main matrix; two-state burst-loss stress profile |
| Bottleneck queue | FIFO tail-drop; one bandwidth-delay product with a 10-packet minimum |

The full design crosses RTT, independent per-direction Bernoulli loss, and MTU at 100 Mbit/s; bandwidth is varied separately, and a burst-loss stress profile is added. The retained TLS summary instead lists configured loss settings of 0%, 1%, 3%, and 5% and provides no emulator commands or achieved-loss distributions. It therefore must not be described as execution of Table 15's principal matrix.

### 7.6 Workloads and Measurements

The complete study would evaluate full establishment, repeated IKE/IPsec rekey, steady-state TCP and UDP transfer, and increasing concurrent establishment load. The current retained results cover only aggregate TLS establishment latency.
Table 16 provides operational definitions for every reported metric.

**Table 16. Measurement definitions**

| Measurement | Operational definition |
|---|---|
| TLS latency | Start of TLS handshake to completion at the initiator |
| IKEv2 latency | First IKE_SA_INIT transmission to installed Child SA |
| Rekey time | Rekey initiation to active replacement SAs |
| Handshake size | Total protocol and transport bytes before secure readiness |
| Processor cost | User time, system time, and cycles per establishment |

| Measurement | Operational definition |
|---|---|
| Memory cost | Peak resident memory and change from baseline |
| Throughput | Application data transferred per second after setup |
| Retransmission | Transport or protocol retransmissions per establishment |
| Fragmentation | IKE fragments, IP fragments, and fragment-related failures |
| Failure | Timeout, incorrect negotiation, or failure to establish required state |
| Capacity | Maximum sustained successful handshakes/s within resource limit |
| Downgrade | Whether prohibited traditional-only operation is rejected and logged |

TLS TCP segmentation is reported separately from IP fragmentation. For IKEv2, IKE-layer fragmentation defined by RFC 7383 is separated from IP-layer fragmentation [28].

### 7.7 Security and Negative Testing

73. Peer offers only the traditional group.
74. Peer uses an unsupported or mismatched draft identifier.
75. Hybrid key share is missing, malformed, or invalid.
76. Required IKE additional-key-exchange step is omitted.
77. Negotiation messages or fragments are dropped or reordered.
78. Peer attempts unauthorized fallback.
79. Gateway restarts during establishment or rekey.
80. Configuration change re-enables a prohibited mode.

When hybrid protection is mandatory, establishment of a traditional-only session is a security failure regardless of performance. The test design follows the principle that downgrade resistance must be demonstrated under active interference, not inferred from a successful preferred-path handshake [26]. Functional tests do not replace cryptographic analysis or formal proof.

### 7.8 Repetition and Experimental Control

The full design randomizes traditional and hybrid runs within each condition and uses independently seeded batches. Under independent Bernoulli trials, a zero-failure one-sided 95% bound satisfies $(1 - p^*)^n \leq 0.05$. For $p^* = 0.001$, $n \geq \ln(0.05)/\ln(0.999) = 2994.23$, so the smallest integer is 2,995 failure-free attempts. The retained aggregate summary does not record n and cannot support a failure-bound claim.
When the complete design is executed, the binomial bound is reported only when conditional independence is credible. Batch-level rates and cluster-aware bootstrap intervals address clustering by path state or fragmentation threshold. None of those uncertainty analyses can be reconstructed from the retained aggregates.

### 7.9 Statistical Analysis

For a measurement $X$ with $X_C > 0$, hybrid percentage overhead is $O_X = 100 \times (X_H - X_C) / X_C$ under the same condition. When $X_C = 0$, only the absolute difference and its confidence interval are reported. Count and failure measures are never converted to a percentage with a zero denominator. Latency is decomposed into local cryptographic cost, protocol round-trip cost, and retransmission or recovery cost.

81. Median, p95, and p99 values with 95% bootstrap confidence intervals.
82. Absolute and percentage differences.
83. Failure proportions with binomial confidence intervals.
84. Quantile regression for interactions among profile, RTT, loss, MTU, bandwidth, and hardware class.
85. Declared significance level and multiple-comparison correction where formal tests are used.

The planned full analysis uses a two-sided significance level of 0.05, Holm correction within each protocol-metric family, and quantile regressions containing profile, hardware class, RTT, loss, MTU, bandwidth, and profile-by-condition interactions. The limited retained summary supports descriptive arithmetic only and is not subjected to formal significance testing.

### 7.10 Framework Evaluation

A future framework evaluation should include at least 15 practitioners spanning WAN engineering, security architecture, PKI, and enterprise risk, applying the rubric to at least 12 anonymized services. Agreement can then be assessed using Krippendorff alpha and a bootstrap interval [39]. No practitioner dataset is claimed in this paper.

Sensitivity analysis varies each QES coefficient by +/-20% with renormalization, each ordinal input by +/-1 level, the timing width u, and band cut points by +/-5 points. It reports band stability, wave changes, Spearman rank correlation, and services sensitive to the assumed CRQC horizon. The purpose is to test a prioritization heuristic, not to estimate compromise probability.

### 7.11 Acceptance Decision

86. Prohibited fallback never succeeds.
87. The upper confidence bound for failure remains within the service error budget.
88. p95 and p99 establishment latency remain within the declared SLO.
89. Processor and memory headroom remain above the required reserve.
90. Rekey completes without unacceptable interruption.
91. No uncontrolled IP fragmentation occurs.
92. Required negotiation logs and downgrade alerts are generated.
93. Rollback returns to the last approved state.

A profile is Conditional when its point estimate is acceptable but its confidence interval crosses the limit. It fails when a security requirement or operational limit is violated.

### 7.12 Reproducibility and Data Handling

A complete reproducibility package should include versions and commits, build options, configurations, emulator parameters and seeds, workload scripts, measurement definitions, anonymized per-trial data, analysis scripts, and artifact hashes. The current limited package instead contains a corrected aggregate TLS summary, an arithmetic checker, author-reported metadata and versions, a limited MTU/MSS note, an evidence inventory, publication guidance, and valid file checksums.

## 8 Supported Limited-Evidence Results

Section 8 reports only quantities supported by the retained artifact or by displayed protocol arithmetic. The artifact does not contain per-trial observations, original packet captures, timestamped TLS/IKE logs, complete command output, or an analysis program that reconstructs percentiles. Consequently, the aggregate TLS values are author-reported and arithmetic-checked, not independently reproducible empirical estimates.

### 8.1 Evidence Boundary

- The retained package contains a corrected TLS aggregate summary, an executable pairwise arithmetic checker, retrospective experiment metadata, author-reported environment versions, a limited GRE MTU/TCP MSS note, an evidence inventory, and valid SHA-256 checksums.
- The package does not prove that a particular TLS or IKE group was negotiated, does not recreate percentiles from raw trials, and does not support sample-count, failure-rate, resource, throughput, retransmission, fragmentation, or downgrade-test conclusions.
- Table 17 separates the retained evidence classes from the conclusions that remain outside the available record.

This boundary prevents protocol-derived constants and author-reported aggregates from being presented as packet-level or independently reproduced measurements.

**Table 17. Retained evidence boundary and permitted interpretation**

| Evidence class | Retained material | Permitted interpretation |
|---|---|---|
| TLS latency | Aggregate p50, p95, and p99 at 0%, 1%, 3%, and 5% configured loss | Author-reported aggregate; pairwise arithmetic checked |
| Protocol size | FIPS 203 ML-KEM-768 sizes plus X25519 share sizes | Protocol-derived calculation; not measured behavior |
| Environment/configuration | Reported device/version/link/RTT notes and MTU/MSS excerpt | Context only; configuration arithmetic checked |
| Raw experimental evidence | No per-trial data, PCAPNG, timestamped logs, complete commands, or original analysis | No protocol or statistical reproducibility claim |

### 8.2 Protocol-Derived Size Baseline

Table 18 uses the ML-KEM-768 encapsulation-key and ciphertext sizes in FIPS 203 and the X25519 share size in the cited hybrid TLS profile. The client share is 1,184 + 32 = 1,216 bytes; the server share is 1,088 + 32 = 1,120 bytes; the combined hybrid key-share data is 2,336 bytes, which is 2,272 bytes greater than the 64-byte classical pair. These are protocol-derived quantities, not observations from the retained experiment.

Table 18. TLS key-share sizes derived from FIPS 203 and the cited hybrid profile

| Component | TLS-C: X25519 | TLS-H: X25519MLKEM768 |
|---|---|---|
| Client key share | 32 bytes | 1,216 bytes (1,184 + 32) |
| Server key share | 32 bytes | 1,120 bytes (1,088 + 32) |
| Combined key-share data | 64 bytes | 2,336 bytes |
| Increment over TLS-C | - | 2,272 bytes |

The 1,216-byte hybrid client key share is below the retained 1,436-byte TCP MSS and therefore does not itself require TCP segmentation. The complete ClientHello can be larger because it contains record and handshake framing, cipher suites, extensions, and other fields. Without an original capture, this paper makes no measured segmentation or fragmentation claim.

### 8.3 Retained Aggregate TLS Latency Summary

The normalized artifact preserves aggregate p50, p95, and p99 values for TLS-C and TLS-H at configured loss settings of 0%, 1%, 3%, and 5%. The original per-trial observations and sample counts are unavailable, so the percentiles cannot be recomputed and no confidence interval, hypothesis test, failure bound, or causal interpretation is reported. The arithmetic checker validates every displayed hybrid-minus-classical difference.

Table 19. Retained author-reported aggregate TLS latency summary (arithmetic checked; sample counts unavailable)

| Profile | Configured loss | p50 | p95 | p99 | p95 delta vs TLS-C |
|---|---|---|---|---|---|
| TLS-C | 0% | 12.00 ms | 12.00 ms | 12.04 ms | Baseline |
| TLS-H | 0% | 12.00 ms | 12.00 ms | 12.06 ms | +0.00 ms |
| TLS-C | 1% | 12.04 ms | 12.08 ms | 12.12 ms | Baseline |
| TLS-H | 1% | 12.11 ms | 12.24 ms | 12.38 ms | +0.16 ms |
| TLS-C | 3% | 12.15 ms | 12.36 ms | 18.00 ms | Baseline |
| TLS-H | 3% | 12.28 ms | 12.74 ms | 18.00 ms | +0.38 ms |
| TLS-C | 5% | 12.22 ms | 12.60 ms | 18.00 ms | Baseline |
| TLS-H | 5% | 12.44 ms | 13.23 ms | 24.00 ms | +0.63 ms |

### 8.4 Retained Testbed and Configuration Context

Table 20 records only the testbed and configuration statements present in the limited package. Device, link, and RTT entries remain author-reported. The GRE and TCP values are arithmetically consistent under the stated basic IPv4 assumptions.

Table 20. Retained testbed and configuration context

| Field | Value | Unit | Retained source | Verification | Arithmetic/check | Manuscript use |
|---|---|---|---|---|---|---|
| Platform/software | ZT-800 OS 4.12; BoringSSL 7f83b162 (abbrev.); strongSwan 5.9.11 | - | Version notes | Author-reported | No command output | Context only |
| WAN / base RTT | 1 Gbps DIA / 6.0 ms | mixed | Metadata note | Author-reported | No trace retained | Context only |
| GRE inner MTU | 1476 | bytes | Configuration note | Arithmetic checked | 1500-20-4 | Conditional design value |
| TCP MSS | 1436 | bytes | Configuration note | Arithmetic checked | 1476-20-20 | Conditional design value |

For basic IPv4 GRE without optional fields, the retained calculation is 1500 - 20 - 4 = 1476 bytes of inner IP MTU. Assuming 20-byte inner IPv4 and TCP headers without options, the corresponding MSS is 1476 - 20 - 20 = 1436 bytes.

These equations are conditional design checks. Optional GRE fields, IP/TCP options, IPsec, VLAN tags, IPv6, or additional encapsulation would change the available MTU or MSS. No retained interface transcript establishes that the reported values were active during the latency measurements.

MSS clamping constrains TCP payload size but does not universally prove the absence of IP fragmentation. IKEv2 fragmentation under RFC 7383 is a separate protocol mechanism and is not evaluated by the retained TLS aggregate summary [28].

### 8.5 Measures Excluded from Empirical Interpretation

Table 21 identifies result categories previously associated with the testbed narrative but unsupported by the retained package. Their numerical values are removed from this submission copy rather than preserved as unverified results.

Table 21. Measures excluded because supporting observations were not retained

| Measure group | Retained data | Evidence basis | Paper treatment | Reason |
|---|---|---|---|---|
| IKE execution/performance | No | No original IKE log, capture, or per-trial data | Results excluded | Negotiation and latency cannot be verified |
| Packets/fragments/retransmissions | No | No valid capture or counter dataset | Results excluded | Wire behavior and denominators are unavailable |
| CPU/memory/capacity/throughput/rekey | No | No synchronized resource or workload observations | Results excluded | Point estimates and uncertainty cannot be reproduced |
| Downgrade/negative tests | No | No original scripts, assertions, logs, or captures | Results excluded | Enforcement outcomes cannot be authenticated |

The exclusions preserve a strict relationship between the paper and its artifact: absence of evidence is reported as unavailable, not replaced by estimates or illustrative logs. These categories remain part of the complete method in Section 7 and are priorities for future data collection.

### 8.6 Artifact Package and Reproducibility Status

Table 22 summarizes the files that can be supplied with this manuscript. The arithmetic checker uses the Python standard library and validates the retained aggregate differences. It does not estimate percentiles, generate missing observations, or validate protocol execution.

**Table 22. Limited evidence artifact contents and validation status**

| Artifact | Retained contents | Validation | Limitation |
|---|---|---|---|
| results_summary_corrected.csv | Eight TLS aggregate rows with corrected deltas | Schema normalized; arithmetic checked | No per-trial observations or n |
| analyze_summary.py | Standard-library pairwise delta checker | All p50/p95/p99 differences pass | Does not recreate percentiles |
| Metadata and versions | Reported device, link, RTT, and component versions | Completeness labels applied | Original command output unavailable |
| Configuration note | GRE MTU and TCP MSS excerpt | Conditional arithmetic passes | Operational state not verified |
| Inventory and SHA256SUMS | Claim-status register and package manifest | All listed file hashes pass | Authenticates package, not missing sources |

All files listed in the package manifest pass SHA-256 verification. This establishes integrity of the limited release package after assembly; it does not retroactively authenticate the missing original experiment sources.

Downgrade rejection remains a mandatory deployment gate in the framework, but no downgrade-test outcome is claimed from the retained package. Classical X25519 is still classically encrypted; the policy concern is loss of the required hybrid component, not a transition to cleartext.

### 8.7 Interpretation

#### 8.7.1 Supported Descriptive Relations

For configured loss levels 0%, 1%, 3%, and 5%, the retained TLS-H minus TLS-C p95 differences are 0.00, 0.16, 0.38, and 0.63 ms. Applying the Section 7.9 overhead formula to the corresponding TLS-C p95 values gives 0.00%, 1.32%, 3.07%, and 5.00%. This increasing descriptive sequence is consistent with a loss-sensitive hybrid overhead hypothesis, but it does not establish statistical significance or causation.

#### 8.7.2 Unsupported Inferences

The p99 differences are 0.02, 0.26, 0.00, and 6.00 ms, which are not monotonic across the retained loss settings. Without per-trial data, sample counts, achieved-loss measurements, or packet sequences, the 5% p99 difference cannot be attributed to segmentation, retransmission, or any other mechanism.

### 8.7.3 Data and Artifact Availability

The retained artifact contains aggregate TLS latency summaries for classical and hybrid profiles at four configured packet-loss settings, author-reported environment versions, and a limited MTU/MSS configuration note. Per-trial observations, original packet captures, timestamped TLS and IKE logs, complete device command output, and the original analysis implementation were not retained. The included program checks aggregate arithmetic only and does not independently reproduce protocol execution.

## 8.8 Correlation with the Migration Framework

The retained summary can inform an exploratory laboratory record, but it cannot satisfy the framework's operational-validation or protocol-verification gates. Under the ECS rubric, an author-reported aggregate without raw verification is not equivalent to automated laboratory measurement or direct independent evidence.

Table 19 supports only the descriptive observation that the recorded p95 difference grows across the four listed loss settings. It does not justify changing a service's residual-exposure state, declaring a profile deployment-prepared, or accepting an SLO.
The protocol-size and MTU/MSS arithmetic identify questions for controlled testing, but they do not demonstrate observed packetization. Consequently, a service remains Open or Transitional until negotiation, fallback enforcement, path behavior, and operational limits are independently evidenced.
No current Section 8 result supports IKE performance, CPU cost, memory consumption, throughput, capacity, rekey behavior, fragmentation, retransmission, failure, or downgrade resistance. Those measures remain explicitly prospective.
This evidence boundary is the core correlation rule: mathematical feasibility, author-reported aggregates, reproducible measurements, and production validation are distinct evidence levels and must not be substituted for one another.

Accordingly, the supported result is a limited, arithmetic-consistent TLS aggregate summary accompanied by a transparent record of unavailable evidence.

# 9 Synthetic Enterprise Worked Example

## 9.1 Enterprise Scenario

Enterprise E is a synthetic multinational research and manufacturing organization connecting 42 branches.

## 9.2 Representative Services

Table 23 defines the six synthetic service classes used in the worked example.

Table 23. Synthetic WAN service classes

| ID | Service | Protocol | Primary concern |
|---|---|---|---|

| ID | Service | Protocol | Primary concern |
|---|---|---|---|
| S1 | Research-data exchange | IKEv2/IPsec | Long-term confidentiality |
| S2 | Manufacturing remote management | IKEv2/IPsec | Authentication and availability |
| S3 | Internet-facing employee portal | TLS 1.3 | Exposure and authentication |
| S4 | Cloud administrative service | TLS 1.3 | Supplier dependency and privilege |
| S5 | Backup replication | IKEv2/IPsec | Long-term confidentiality and throughput |
| S6 | Branch collaboration | Overlay and TLS | Scale and equipment lifecycle |

The shared certificate authorities, WAN management platform, and branch software image receive high concentration scores because they affect many services.
Table 24 makes every synthetic QES input and inventory mapping explicit.

Table 24. Synthetic service coverage and QES inputs

| ID | Coverage | L_C/M_C | L_A/M_A | I_C/I_A | V_K/V_A | E_C/E_A | D |
|---|---|---|---|---|---|---|---|
| S1 | 18 IPsec | 15/3 | 8/5 | 4/4 | 5/5 | 4/4 | 3 |
| S2 | 12 IPsec | 6/6 | 8/7 | 4/5 | 5/5 | 1/2 | 4 |
| S3 | 22 TLS | 6/5 | 5/6 | 1/3 | 5/5 | 5/5 | 3 |
| S4 | 9 TLS | 3/6 | 6/7 | 5/5 | 5/5 | 4/3 | 4 |
| S5 | 14 IPsec | 8/4 | 6/5 | 5/4 | 5/5 | 3/3 | 4 |
| S6 | 120 IPsec; 43 TLS | 5/6 | 10/5 | 1/1 | 3/3 | 2/2 | 2 |

Table 25 makes the MRS and ECS vectors reproducible.

Table 25. Synthetic readiness and evidence vectors

| ID | MRS vector: inventory, support, interoperability, validation, governance | ECS vector: lifetime, lead time, impact, control, path, dependency, readiness, verification |
|---|---|---|
| S1 | 3,2,1,1,2 | 4,3,4,3,4,3,3,3 |
| S2 | 3,2,2,1,3 | 3,3,3,3,3,3,3,3 |
| S3 | 4,4,4,4,4 | 5,4,5,4,5,4,4,5 |
| S4 | 2,1,1,1,2 | 3,2,3,2,3,2,2,2 |
| S5 | 4,3,3,3,4 | 4,4,4,4,5,4,4,4 |
| S6 | 3,3,3,3,3 | 3,3,4,4,4,4,4,4 |

### 9.3 Initial Assessment

Table 26 summarizes the Q = 10 prioritization decision.

Table 26. Synthetic service priority and preparation summary

| Service | QES_C | QES_A | QES priority | MRS | ECS | Decision |
|---|---|---|---|---|---|---|
| S1 Research | 87 | 81 | 87 Critical | 36 Low | 68 Moderate | Wave 0 |
| S2 Manufacturing | 75 | 91 | 91 Critical | 44 Low | 60 Moderate | Wave 0 |

| Service | QES_C | QES_A | QES priority | MRS | ECS | Decision |
|---|---|---|---|---|---|---|
| S3 Portal | 62 | 72 | 72 High | 80 Deployment-prepared | 90 High | Wave 1 |
| S4 Cloud admin | 77 | 88 | 88 Critical | 28 Low | 48 Low | Wave 0 |
| S5 Backup | 84 | 77 | 84 Critical | 68 Partial | 82 High | Wave 0 |
| S6 Collaboration | 45 | 53 | 53 Moderate | 60 Partial | 75 Moderate | Wave 2 |

For S1, L_C = 15 years, M_C = 3 years, and Q = 10 years yield H_C = 8 and T_C = 5. With I_C = 4, V_K = 5, E_C = 4, and D = 3, QES_C = 87. Readiness inputs 3, 2, 1, 1, and 2 produce MRS = 36. Evidence grades 4, 3, 4, 3, 4, 3, 3, and 3 produce ECS = 20(0.15(4) + 0.10(3) + 0.15(4) + 0.15(3) + 0.10(4) + 0.10(3) + 0.15(3) + 0.10(3)) = 68. The result is immediate treatment and an urgent pilot, not an untested production change.
Table 27 shows which service bands depend on the assumed Q horizon.

**Table 27. Synthetic QES scenario sensitivity**

| ID | QES_5 | QES_10 | QES_15 | Interpretation |
|---|---|---|---|---|
| S1 | 87 Critical | 87 Critical | 81 Critical | Band stable across scenarios |
| S2 | 91 Critical | 91 Critical | 78.5 High | Band changes with Q; review timing assumptions |
| S3 | 82 Critical | 72 High | 59.5 Moderate | Band changes with Q; review timing assumptions |
| S4 | 93 Critical | 88 Critical | 75.5 High | Band changes with Q; review timing assumptions |
| S5 | 93 Critical | 84 Critical | 69 High | Band changes with Q; review timing assumptions |
| S6 | 57 Moderate | 53 Moderate | 40.5 Moderate | Band stable across scenarios |

### 9.4 Interpretation and Treatment

S1 ranks highly because recorded research data may retain value beyond the assumed quantum horizon. S2 is driven primarily by future authentication and availability consequences. S3 is publicly exposed but comparatively ready, making it an accelerated migration candidate. S4 combines privileged access, supplier dependence, and weak evidence. S5 requires path, throughput, and rekey validation. S6 can follow a planned hardware lifecycle if procurement requirements are enforced.
Table 28 connects each synthetic service to treatment and required evidence.

**Table 28. Synthetic treatment plan**

| Service | Immediate treatment | Required evidence |
|---|---|---|
| S1 | Reduce exposure; evaluate an enforced interim control; begin hybrid pilot | Interoperability, MTU, downgrade, and key management |
| S2 | Restrict management paths; test outside production control windows | Availability, restart, authentication, and rollback |
| S3 | Controlled hybrid TLS rollout to managed clients | Tail latency, failure, negotiation logs, and fallback rejection |
| S4 | Escalate supplier; restrict administration; create expiring exception | Profile detail, independent evidence, and delivery dates |
| S5 | Pilot hybrid IKEv2 on a representative replication path | Throughput, fragmentation, rekey, CPU, and recovery |

| Service | Immediate treatment | Required evidence |
|---|---|---|
| S6 | Include post-quantum requirements in gateway refresh | Procurement, logging, configuration, and lifecycle commitments |

### 9.5 Dependency-Aware Sequence

8. Complete inventory for the certificate authorities and management platform.
9. Define approved profiles and enable negotiation and downgrade visibility.
10. Validate profiles in the laboratory.
11. Migrate research and backup paths after gates pass.
12. Migrate manufacturing management during approved operational windows.
13. Expand managed-client TLS and replace branch gateways.
14. Migrate remaining authentication dependencies and disable prohibited traditional-only operation.

### 9.6 Conditional Residual-Exposure Example

If S5 passes every confidentiality gate, enforces the approved hybrid profile, independently verifies negotiation, and rejects prohibited fallback, its confidentiality residual-exposure state moves from Open to Prospectively closed. The QES priority record remains available to show why the service was migrated early, but no point subtraction is reported. S5 remains Transitional at the service level until its certificate, identity, and trust dependencies also satisfy the authentication gates. This is a conditional worked example, not an observed production transition.

### 9.7 Lessons from the Worked Example

94. Protection lifetime can outweigh public exposure when recorded data remains valuable for many years.
95. Authentication risk can determine overall priority for management and control services.
96. Low readiness and weak evidence increase program urgency rather than reducing exposure.
97. A deployment-prepared High-priority service may be deployed before a low-readiness Critical service, while Critical treatment continues in parallel.
98. Post-quantum key establishment does not complete migration while identity and trust dependencies remain vulnerable.

## 10 Discussion and Limitations

### 10.1 Main Contribution

The main contribution is the connection between cryptographic exposure, operational evidence, and migration action. Existing guidance commonly treats inventory, risk, protocol testing, procurement, and rollout as related but separate activities. The proposed framework links them through one service-level record and an auditable evidence package.

A single composite could let strong supplier preparation mask severe exposure or let high theoretical exposure hide the absence of deployable controls. Keeping QES, MRS, and ECS separate exposes those tensions to technical and business owners.

### 10.2 Practical Security Implications

The framework shows why enterprise quantum risk cannot be reduced to an algorithm inventory. Confidentiality urgency depends on how long information must remain protected and how long migration will take. Authentication risk depends on the lifetime and concentration of identities, trust anchors, signed software, and management systems. Path behavior determines whether a correct cryptographic design can operate reliably across a real WAN.

Hybrid key establishment is useful during coexistence, but it is not automatically an end state. Its value depends on correct composition, independent component security, enforceable negotiation, and operational evidence. A service should not be called quantum-ready solely because a product interface displays an ML-KEM option.

### 10.3 Policy and Governance Implications

The framework operationalizes the recurring program established in Section 2.6: discovery, data and system prioritization, dependency mapping, accountable planning, and validation [7, 15, 18, 19, 37]. It does not replace legal, regulatory, or sector-specific analysis, and use of the model is not a certification of compliance.

For private enterprises, government directives can inform governance even when they are not directly binding. The service record, evidence package, exception expiry, and residual-risk decision provide artifacts that can be mapped to an organization's existing risk, audit, procurement, and change-management processes.

### 10.4 Interpreting the Limited Performance Evidence

The retained Table 19 values are descriptive aggregates. Their p95 absolute differences increase from 0.00 to 0.63 ms across the listed loss settings, and the corresponding relative differences increase from 0.00% to 5.00%. Because n, per-trial values, achieved loss, timing boundaries, and uncertainty intervals are unavailable, this pattern cannot establish statistical significance, implementation causality, or service-level acceptability.

Negative security tests retain priority over favorable averages, but no negative-test outcome is claimed in this version. A profile that permits prohibited fallback would fail the security gate regardless of latency. Same-stack success would also provide weaker evidence than cross-stack interoperability, and a laboratory pass would not establish production readiness without monitoring, rollback, and ownership.

### 10.5 Limitations and Threats to Validity

**Quantum-horizon uncertainty.** Q is a scenario input, not a forecast. Scores may change when the assumed horizon changes. Reporting QES_5, QES_10, and QES_15 makes this dependence visible but does not remove uncertainty.

**Scoring judgement.** Protection lifetime, impact, exposure, and dependency concentration require practitioner judgement. The rubrics improve consistency but cannot eliminate organizational bias. Expert review, evidence grading, and inter-rater analysis are therefore required.
Ordinal aggregation. I, V, E, and D are ordered judgement scales, and their adjacent levels have not been shown to be equally spaced. QES is therefore limited to ranking and banding under a common rubric. Point subtraction, probability language, and claims about the amount of risk removed are excluded pending practitioner validation [38].
Closure semantics. A high-impact service can retain a high intrinsic priority even after migration. The framework therefore uses gate-backed residual-exposure states, rather than a lower QES alone, to represent prospective closure and quantum-ready status.
**Default weights.** The proposed weights express an initial decision model and have not been established as universal empirical constants. Sensitivity analysis can identify unstable cases, but validation across sectors remains necessary.
Protocol scope. The supported observational data cover only aggregate TLS-C and TLS-H latency values. The complete design includes IKEv2/IPsec, but no IKE execution result is claimed. Other VPN, SD-WAN, routing, SSH, remote-access, certificate, and software-signing mechanisms are outside the current evidence.
Authentication scope. The proposed performance comparison holds traditional authentication constant. The retained data do not validate certificate behavior, post-quantum signatures, identity migration, or end-to-end authentication.
Protocol maturity. The TLS hybrid and IKEv2 ML-KEM specifications cited in [8] and [9] remain Internet-Drafts and works in progress as of August 10, 2026. Any future experiment must identify the exact revision, implementation commit, and collection date and must be repeated after material protocol changes.
Implementation and hardware effects. The retained environment versions are author-reported and lack build or command-output verification. Cryptographic libraries, compiler options, acceleration, operating systems, gateway code, and processor architectures may produce different results.
Path realism. The retained notes identify a 1 Gbps DIA service, a reported 6.0-ms RTT, and conditional MTU/MSS values, but no raw path measurements or impairment commands were retained. The summary therefore cannot establish representativeness for other carrier, wireless, middlebox, congestion, or routing conditions.

### 10.6 Future Work

99. Execute the complete TLS and IKEv2 design with per-trial records, valid captures, timestamped logs, exact builds, independently achieved impairment measurements, and reproducible analysis.
100. Evaluate ML-DSA and SLH-DSA certificate chains, trust-anchor migration, enrollment, and revocation.
101. Validate QES, MRS, and ECS with practitioners from multiple sectors and jurisdictions.
102. Extend the model to remote access, SD-WAN controllers, routing security, SSH, software signing, and constrained devices.
103. Measure migration cost, staffing, supplier lead time, energy use, and hardware-replacement effects.
104. Automate inventory and dependency updates while preserving evidence provenance and human risk ownership.

105. Study long-term drift, fallback events, and operational incidents after production deployment.

## 11 Conclusion

Enterprise WANs face a dual transition problem. Recorded traffic protected by RSA- or elliptic-curve-based key establishment may be exposed in the future, while authentication, trust, and management dependencies may become vulnerable when a CRQC is available. Migration must address these threats without creating unacceptable outages, fragmentation, interoperability failures, or silent downgrade.

This paper presents a vendor-neutral framework that connects a WAN threat model to four separate decision outputs: confidentiality exposure priority, authentication exposure priority, migration preparation, and evidence confidence. These outputs guide migration waves, control selection, laboratory and pilot gates, procurement, exception handling, and categorical residual-exposure decisions. The framework is intended to translate standards and government direction into service-level action without treating an ordinal score as a probability or amount of risk.

The experimental design defines how classical and X25519/Curve25519 plus ML-KEM-768 profiles should be compared across TLS 1.3 and IKEv2/IPsec. The retained artifact supports only an arithmetic-checked aggregate TLS latency summary and protocol-derived size and MTU/MSS calculations. It does not support claims about IKE execution, raw packet behavior, resources, throughput, failure, rekey, or downgrade resistance. The paper should therefore be evaluated as a vendor-neutral migration framework with limited descriptive evidence, not as validation that a particular implementation or enterprise WAN is quantum-ready.

## References

[1] P. W. Shor. 1997. Polynomial-Time Algorithms for Prime Factorization and Discrete Logarithms on a Quantum Computer. SIAM Journal on Computing 26, 5, 1484-1509. https://doi.org/10.1137/S0097539795293172

[2] D. Joseph, R. Misoczki, M. Manzano, et al. 2022. Transitioning Organizations to Post-Quantum Cryptography. Nature 605, 237-243. https://doi.org/10.1038/s41586-022-04623-2

[3] National Institute of Standards and Technology. 2024. Module-Lattice-Based Key-Encapsulation Mechanism Standard. FIPS 203. https://doi.org/10.6028/NIST.FIPS.203

[4] National Institute of Standards and Technology. 2024. Module-Lattice-Based Digital Signature Standard. FIPS 204. https://doi.org/10.6028/NIST.FIPS.204

[5] National Institute of Standards and Technology. 2024. Stateless Hash-Based Digital Signature Standard. FIPS 205. https://doi.org/10.6028/NIST.FIPS.205

[6] Office of Management and Budget. 2022. Migrating to Post-Quantum Cryptography. Memorandum M-23-02. https://www.whitehouse.gov/wp-content/uploads/2022/11/M-23-02-M-Memo-on-Migrating-to-Post-Quantum-Cryptography.pdf

[7] Office of Management and Budget. 2026. Execution of the Migration to Post-Quantum Cryptography. Memorandum M-26-15. https://www.whitehouse.gov/wp-content/uploads/2026/06/M-26-15-Execution-of-the-Migration-to-Post-Quantum-Cryptography.pdf

[8] K. Kwiatkowski, P. Kampanakis, B. Westerbaan, and D. Stebila. 2026. Post-Quantum Hybrid ECDHE-MLKEM Key Agreement for TLS 1.3. IETF Internet-Draft draft-ietf-tls-ecdhe-mlkem-05, work in progress. https://datatracker.ietf.org/doc/draft-ietf-tls-ecdhe-mlkem/05/

[9] P. Kampanakis. 2026. Post-Quantum Key Exchange with ML-KEM in the Internet Key Exchange Protocol Version 2 (IKEv2). IETF Internet-Draft draft-ietf-ipsecme-ikev2-mlkem-05, work in progress. https://datatracker.ietf.org/doc/draft-ietf-ipsecme-ikev2-mlkem/05/

[10] D. Ott, C. Peikert, B. Hesse, et al. 2019. Identifying Research Challenges in Post Quantum Cryptography Migration and Cryptographic Agility. arXiv:1909.07353. https://doi.org/10.48550/arXiv.1909.07353

[11] L. K. Grover. 1996. A Fast Quantum Mechanical Algorithm for Database Search. In Proceedings of the Twenty-Eighth Annual ACM Symposium on Theory of Computing, 212-219. https://doi.org/10.1145/237814.237866

[12] European Telecommunications Standards Institute. 2020. Quantum-Safe Hybrid Key Exchanges. ETSI TS 103 744 V1.1.1. https://www.etsi.org/deliver/etsi_ts/103700_103799/103744/01.01.01_60/ts_103744v010101p.pdf

[13] C. J. Tjhai, M. Tomlinson, G. Bartlett, S. Fluhrer, D. Van Geest, O. Garcia-Morchon, and V. Smyslov. 2023. Multiple Key Exchanges in IKEv2. RFC 9370. https://doi.org/10.17487/RFC9370

[14] D. Moody, R. Perlner, A. Regenscheid, A. Robinson, and D. Cooper. 2024. Transition to Post-Quantum Cryptography Standards. NIST IR 8547, Initial Public Draft. https://doi.org/10.6028/NIST.IR.8547.ipd

[15] UK National Cyber Security Centre. 2025. Timelines for Migration to Post-Quantum Cryptography. https://www.ncsc.gov.uk/guidance/pqc-migration-timelines

[16] Association for Computing Machinery. 2026. Digital Threats: Research and Practice: Journal Scope and Author Information. Accessed August 9, 2026. https://dl.acm.org/journal/dtrap

[17] K. F. Hasan, L. Simpson, M. A. R. Baee, C. Islam, Z. Rahman, W. Armstrong, P. Gauravaram, and M. McKague. 2024. A Framework for Migrating to Post-Quantum Cryptography: Security Dependency Analysis and Case Studies. IEEE Access 12, 23427-23450. https://doi.org/10.1109/ACCESS.2024.3360412

[18] W. Newhouse, M. Souppaya, W. Barker, and C. Brown. 2023. Migration to Post-Quantum Cryptography: Quantum Readiness: Cryptographic Discovery. NIST SP 1800-38B, Preliminary Draft. https://csrc.nist.gov/pubs/sp/1800/38/iprd-%281%29

[19] National Institute of Standards and Technology. 2026. Considerations for Achieving Crypto Agility: Strategies and Practices. NIST CSWP 39upd1, final update June 29, 2026. https://doi.org/10.6028/NIST.CSWP.39-upd1

[20] D. Sikeridis, P. Kampanakis, and M. Devetsikiotis. 2020. Post-Quantum Authentication in TLS 1.3: A Performance Study. In Network and Distributed System Security Symposium. https://doi.org/10.14722/ndss.2020.24203

[21] M. Sosnowski, F. Wiedner, E. Hauser, L. Steger, D. Schoinianakis, S. Gallenmuller, and G. Carle. 2023. The Performance of Post-Quantum TLS 1.3. In CoNEXT Companion 2023, 19-27. https://doi.org/10.1145/3624354.3630585

[22] S. Bae, Y. Chang, H. Park, M. Kim, and Y. Shin. 2023. A Performance Evaluation of IPsec with Post-Quantum Cryptography. In Information Security and Cryptology - ICISC 2022, LNCS 13849, 249-266. https://doi.org/10.1007/978-3-031-29371-9_13

[23] A. Mutlugun, Y. Hanna, and K. Akkaya. 2024. Performance Evaluation of Quantum-Resistant IKEv2 Protocol for Satellite Networking Environments. In 2024 IEEE Virtual Conference on Communications, 1-7. https://doi.org/10.1109/VCC63113.2024.10914395

[24] G. Twardokus, W. Joslin, B. Carini, H. Rahbari, and W. Layton. 2024. Internet Measurement of Quantum-Resistant IKEv2 in Constrained Networks. arXiv:2411.15936, preprint. https://doi.org/10.48550/arXiv.2411.15936

[25] J. A. Montenegro, R. Rios, and J. Lopez-Cerezo. 2026. A Performance Evaluation Framework for Post-Quantum TLS. Future Generation Computer Systems 175, 108062. https://doi.org/10.1016/j.future.2025.108062

[26] K. Bhargavan, C. Brzuska, C. Fournet, M. Green, M. Kohlweiss, and S. Zanella-Beguelin. 2016. Downgrade Resilience in Key-Exchange Protocols. In 2016 IEEE Symposium on Security and Privacy. https://doi.org/10.1109/SP.2016.37

[27] National Institute of Standards and Technology. 2025. Recommendations for Key-Encapsulation Mechanisms. NIST SP 800-227, Final. https://doi.org/10.6028/NIST.SP.800-227

[28] V. Smyslov. 2014. Internet Key Exchange Protocol Version 2 (IKEv2) Message Fragmentation. RFC 7383. https://doi.org/10.17487/RFC7383

[29] Joint Task Force Transformation Initiative. 2012. Guide for Conducting Risk Assessments. NIST SP 800-30 Rev. 1. https://doi.org/10.6028/NIST.SP.800-30r1

[30] National Institute of Standards and Technology. 2004. Standards for Security Categorization of Federal Information and Information Systems. FIPS 199. https://doi.org/10.6028/NIST.FIPS.199

[31] S. Fluhrer, P. Kampanakis, D. McGrew, and V. Smyslov. 2020. Mixing Preshared Keys in IKEv2 for Post-Quantum Security. RFC 8784. https://doi.org/10.17487/RFC8784

[32] V. Smyslov. 2025. Mixing Preshared Keys in the IKE_INTERMEDIATE and CREATE_CHILD_SA Exchanges of IKEv2 for Post-Quantum Security. RFC 9867. https://doi.org/10.17487/RFC9867

[33] E. Rescorla. 2018. The Transport Layer Security (TLS) Protocol Version 1.3. RFC 8446. https://doi.org/10.17487/RFC8446

[34] C. Kaufman, P. Hoffman, Y. Nir, P. Eronen, and T. Kivinen. 2014. Internet Key Exchange Protocol Version 2 (IKEv2). RFC 7296. https://doi.org/10.17487/RFC7296

[35] Y. Nir, S. Josefsson, and M. Pegourie-Gonnard. 2016. Curve25519 and Curve448 for IKEv2. RFC 8031. https://doi.org/10.17487/RFC8031

[36] V. Smyslov. 2022. Intermediate Exchange in the Internet Key Exchange Protocol Version 2 (IKEv2). RFC 9242. https://doi.org/10.17487/RFC9242

[37] Executive Office of the President. 2026. Securing the Nation Against Advanced Cryptographic Attacks. Executive Order 14412, June 22, 2026. https://www.whitehouse.gov/presidential-actions/2026/06/securing-the-nation-against-advanced-cryptographic-attacks/

[38] J. Spring, E. Hatleback, A. D. Householder, A. Manion, and D. Shick. 2021. Time to Change the CVSS? IEEE Security & Privacy 19, 2, 74-78. https://doi.org/10.1109/MSEC.2020.3044475

[39] K. Krippendorff. 2018. Content Analysis: An Introduction to Its Methodology, 4th ed. SAGE Publications.